\documentclass[3p,times,twocolumn]{elsarticle}

\usepackage{ecrc}
\usepackage{epsfig}

\volume{00}

\firstpage{1}

\journalname{Nuclear Physics B Proceedings Supplement}

\runauth{A. Nucciotti}

\jid{nuphbp}

\jnltitlelogo{Nuclear Physics B Proceedings Supplement}

\usepackage{amssymb}
\usepackage[figuresright]{rotating}

\def\Journal#1#2#3#4{{#1} {\bf #2}(#3), p. #4.}

\def\NIMA{Nucl.\,Instr. and Meth. A}

\def\NPB{Nucl.\,Phys. B}
\def\NDS{Nucl. Data Sheets}
\def\APP{Astropart. Phys.}
\def\NPA{Nucl.\,Phys. A}

\def\PLB{Phys.\,Lett.  B}

\def\JPGNP{J.\,Phys. G: Nucl. Phys.}
\def\PRL{Phys.\,Rev.\,Lett.}
\def\PRA{Phys.\,Rev. A}

\def\PRC{Phys.\,Rev. C}

\def\JLT{J.\,Low\,Temp.\,Phys.}

\def\Re{$^{187}$Re}
\def\Ho{$^{163}$Ho}

\def\Tr{$^3$H}
\def\mn{$m_\nu$}
\def\mug{$\mu$g}

\def\mus{$\mu$s}

\def\de{$\Delta E$}

\def\b{$\beta$}

\def\agre{AgReO$_4$}

\begin{document}

\begin{frontmatter}

\dochead{}
%% Use \dochead if there is an article header, e.g. \dochead{Short communication}

\title{Neutrino mass calorimetric searches in the MARE experiment}

\author[mib]{A.\,Nucciotti\ead{angelo.nucciotti@mib.infn.it}}
\author{for the MARE Collaboration}

\address[mib]{Universit\`a di Milano-Bicocca e INFN Sezione di Milano-Bicocca, Milano, Italia}

\begin{abstract}
%% Text of abstract
The international project ``Microcalorimeter Arrays for a Rhenium Experiment'' (MARE) aims at the direct and calorimetric measurement of the electron neutrino mass with sub-electronvolt sensitivity. Calorimetric neutrino mass experiments measure all the energy released in a beta decay except for the energy carried away by the neutrino, therefore removing the most severe systematic uncertainties which have plagued the traditional and, so far, more sensitive spectrometers. Calorimetric measurements are best realized exploiting the thermal detection technique. This approach uses thermal microcalorimeters whose absorbers contain a low transition energy $Q$ beta decaying isotope. To date the two best options are \Re\ and \Ho. While the first beta decays, the latter decays via electron capture, but both have a $Q$ value around 2.5 keV. The potential of using \Re\ for a calorimetric neutrino mass experiment has been already demonstrated. On the contrary, no calorimetric spectrum of \Ho\ has been so far measured with the precision required to set a useful limit on the neutrino mass. In this talk we present the status and the perspectives of the MARE project activities for the active isotope selection and the single channel development. We also discuss the neutrino mass statistical sensitivity achievable with both isotopes.
\end{abstract}

\begin{keyword}
%% keywords here, in the form: keyword \sep keyword

%% MSC codes here, in the form: \MSC code \sep code
%% or \MSC[2008] code \sep code (2000 is the default)
neutrino mass \sep beta decay \sep electron capture \sep low temperature detectors
\end{keyword}

\end{frontmatter}

%%
%% Start line numbering here if you want
%%
% \linenumbers

%% main text
%
\section{Introduction}
\label{sec:intro}
The  ``Microcalorimeter Arrays for a Rhenium Experiment'' (MARE) project aims at providing a competitive alternative to the KATRIN experiment in 
the effort of measuring the neutrino mass \mn\ with a sensitivity as low as 0.1\,eV.

To date, the most sensitive experiments were carried out analyzing the \Tr\ decay in magnetic adiabatic collimation spectrometers with electrostatic filter, yielding an upper limit on the electron anti-neutrino mass of
2.2\,eV \cite{MainzTroitsk}. The soon starting experiment KATRIN will analyze the \Tr\ beta decay end-point  with a
much more sensitive electrostatic spectrometer and with an expected  statistical sensitivity of about 0.2\,eV \cite{KATRIN}.

However, these experiments suffer from many systematic uncertainties since the \Tr\ source is external to the spectrometer and the measured electron energy has to be corrected for the energy lost in exciting atomic and molecular states, 
in crossing the source, in scattering through the spectrometer, and more. 
Therefore, to improve the sensitivity, it is necessary to reduce both the systematic and the statistical uncertainties. 

Because of the large weight of systematics, it is inherent in neutrino mass measurements
that confidence in the results can be obtained only through confirmation by independent
experiments. 

\section{The MARE project}
\label{sec:mare}
An alternative approach to spectrometry is calorimetry where the beta source is embedded in the detector so that all the energy emitted in the decay is measured, except the one taken away by the neutrino.
In this configuration, the systematic uncertainties arising from the external electron source are eliminated. 
On the other hand, since calorimeters detect all the decays occurring over the entire beta energy spectrum, the source activity must be limited to avoid spectral distortions and background at the end-point due to pulse pile-up.
As a consequence the statistics near the end-point is limited as well.

Direct kinematical neutrino mass experiments look for a deformation in the energy spectrum of visible particles for neutrino
kinetic energies in an interval $\Delta E$ approaching zero -- i.e. at the spectrum end-point $Q$. The fraction of these events in beta decay is only $\propto (\Delta E/Q)^3$. The limitation to the statistics may be then partially balanced by using 
beta decaying isotopes with an end-point energy as low as possible, like \Re, which has the second lowest known transition energy ($Q\sim2.5$\,keV).

A perfect practical way to make a calorimetric measurement is to use thermal detectors. 
At thermal equilibrium, the temperature rise of the detector -- measured by a suitable thermometer -- is due to  the sum of the energy of the emitted electron and of
all other initial excitations. 
The measurement is then free from the systematics induced by any possible energy
loss in the source and is not affected by problems related to decays on excited final states.
In principle one remaining systematics may be due to energy lost in metastable 
states living longer than the detector integration time which is always more than about 1\,\mus.

The Microcalorimeter Arrays for a Neutrino Mass Experiment (MARE) project was launched 
by a large international collaboration in 2005 with the aim of measuring the neutrino mass with a calorimetric approach.
The baseline of the MARE project consists in a large array of rhenium based thermal detectors. 
As discussed in the following section, different options for the isotope are also being considered. 

\subsection{Two isotopes: $^{187}$Re vs.  $^{163}$Ho option}
\label{sec:isotopes}
Rhenium is in principle perfectly suited for fabricating thermal detectors. Metallic rhenium crystals undergo a superconducting transition below 1.6\,K allowing to reach very high sensitivity thanks to the reduced thermal capacity
in the superconducting state. Moreover, the lifetime -- about $4.5\times10^9$\,years -- and the natural isotopic abundance
of \Re\ -- about 60\% -- give an ideal decay rate of about 1\,decay per second per milligram. Also dielectric compounds in
proper crystalline forms can be used.
To date, only two experiments have been carried out with thermal detectors containing \Re: 
the MANU \cite{MANU-PRC,MANU} and MIBETA \cite{MIBETA-PRL,MIBETA-LTD10} experiments.
MANU used one detector with a NTD thermistor glued to a 1.6\,mg metallic rhenium single crystal,
while MIBETA used an array of eight silicon implanted thermistors with \agre\ crystals for a total mass of about 2.2\,mg.
Both the experiments collected a statistics of about 10$^7$ decays, yielding
limits on \mn\ of about 26\,eV at 95\% CL and 15\,eV at 90\% CL, respectively. The systematics affecting these experiments are still small compared with the statistical error. 

Lowering these limits requires better performing detectors and larger statistics. From a simple statistical analysis for the 
90\% CL limit on \mn,  $\Sigma_{90}(m_\nu)$, one can write \cite{Nucciotti-Sens} 
\begin{equation}
\label{eq:sensitivity}
\Sigma_{90}(m_\nu) \propto  \frac{Q}{\sqrt[4]{N_{ev}}} \left[{\frac{ \Delta E}{Q} + \frac{Q}{\Delta E} \left( \frac{3}{10} \ f_{pp}
+b \frac{Q}{A_\beta}\right)}\right]^{\frac{1}{4}}
\end{equation}
where: i) $A_\beta$ is the \b\ activity of each detector. ii) $b$ is the average background rate for unit energy and for a single detector. iii) $N_{ev}=A_\beta N_{det} t_M$ is the total number of events, with $N_{det}$ and $t_M$ the number of detectors and the measuring time, respectively.  
iv) $f_{pp}$ is the probability that two \b\ decays pile-up and it is given by $\tau A_\beta$, where $\tau$ is the time resolution, strictly related to the detector rise time.
Two decays happening closer than $\tau$ are mistaken as one single decay with energy equal to
the sum of the two decay energies. The frequency of these piling up events is about $\tau A_\beta^2$ and their spectrum
$N_{pile-up}(E) \propto N(E)\bigotimes N(E)$ -- where $N(E)$ is the \Re\ \b\ spectrum -- extends from 0 to about $2 Q$.
v) Finally $\Delta E$ is the optimal analysis interval given by the largest between the detector energy resolution $\Delta E_\mathrm{FWHM}$ and $Q (0.3 f_{pp}+bQ/A_\beta)^{1/2}$ (see \cite{Nucciotti-Sens} for more details).
 
The three terms in eq.\,(\ref{eq:sensitivity}) arise respectively from the statistical fluctuations of the \b, pile-up and background spectra. Eq.\,(\ref{eq:sensitivity}) shows the importance of improving the detector energy resolution and of minimizing pile-up by reducing the detector rise time. On the other hand it shows also that the largest reduction on the limit can only come by substantially increasing the total statistics $N_{ev}$. If pile-up and background are kept negligible the sensitivity improves as $\sqrt[4]{1/N_{ev}}$. Therefore improving the present limit of MIBETA and MANU
by a factor 100,  would require to increase the statistics by a factor $10^8$, i.e. to collect a statistics of
about $10^{15}$ decays. 

A more accurate estimate of the sensitivity can be obtained by Montecarlo frequentist approach \cite{Nucciotti-Sens}.  As discussed in \cite{Nucciotti-Sens}, the same Montecarlo approach may be used also to investigate the source of
systematic uncertainties peculiar to the calorimetric technique: from this study it appears that the most crucial 
sources are the Beta Environmental Fine Structure (BEFS) \cite{Ge-Nature,BEFS-PRL}, the theoretical spectral shape of the \Re\ \b\ decay, the background and the pile-up \cite{Foggetta-LTD12}.

\begin{figure}
\begin{center}
 \includegraphics*[bb=40 20 580 430,clip,width=\linewidth]{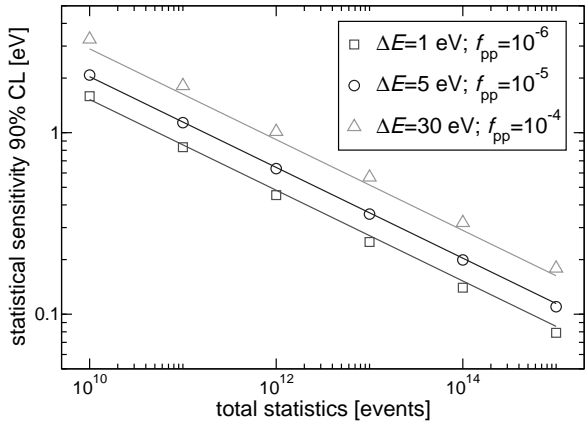}
\caption{\label{fig:sens} Statistical sensitivity of a \Re\ neutrino mass calorimetric experiment estimated with both Montecarlo (points) and analytic (curves) approaches, assuming $b=0.$}
\end{center}
\end{figure}
Considering only the statistical sensitivity, Fig.\,\ref{fig:sens} shows the results of both the analytic and the Montecarlo approaches  for few hypothetical experimental parameters.
The first two rows of Tab.\,\ref{tab:sens} show the total statistics and exposures $T=N_{det}t_M$ required to reach a statistical sensitivity  $\Sigma_{90}(m_\nu)$ of 0.2 and 0.1\,eV, respectively, for two plausible experimental configurations.
\begin{table*}[ht!]
\caption{\label{tab:sens} Experimental exposure required for the target statistical sensitivity in the second column, with $b=0$ and $Q_{EC}=2.2$\,keV.}
\begin{center}
\begin{tabular}{ccccccc}
\hline 
isotope & sensitivity & $A_\beta$ &	$\tau$ &	\de\ &	$N_{ev}$ &	exposure $T$ \\[0pt] % 
& [eV] & [Hz] & 	[$\mu$s] &	[eV] &	[counts] &	[detector$\times$year] \\
\hline 
\Re\ & 0.2 & 10&	3&	3&	$1.3\times10^{14}$&	$4.1\times10^{5}$\\
\Re\ & 0.1 & 10&	1&	1&	$10.3\times10^{14}$&	$3.3\times10^{6}$\\
\hline 
\Ho\ & 0.2 & 100&	0.1&	1&	$4.6\times10^{13}$&	$0.15\times10^{5}$\\
\Ho\ & 0.1 & 100&	0.1&	0.3&	$6.4\times10^{14}$&	$0.20\times10^{6}$\\
\hline 
\end{tabular} 
\end{center}
\end{table*}
For example a target neutrino mass statistical sensitivity of 0.2\,eV could be expected running for 10\,years $4\times10^4$ rhenium detectors, each with a mass of 10\,mg -- for an activity of about 10\,Hz --  and with energy and time resolutions of about 3\,eV and 3\,\mus\ respectively. The total required mass of natural rhenium is about 400\,g.
This calls for the use of large arrays  -- for a total number of channels of the order of 10000 -- coupled to an appropriate signal multiplexing scheme to avoid running into insurmountable cryogenic and economic problems.
Altogether these requirements are about the same as for IXO -- the next generation X-ray space observatory -- and largely proved to be technically feasible. The major open issue remains the  metallic rhenium absorber coupling to the sensor.
In fact, in spite of the many efforts, persisting technical difficulties prevented from realizing the target performances with rhenium absorbers so far.
In order to have a viable alternative to the baseline MARE design using rhenium \b\ decay, the MARE collaboration is
considering the possibility to use \Ho\ electron capture \cite{Gatti-LTD12}.

 Since the '80s \Ho\ electron capture (EC) decay has been the subject of many experimental investigations as a powerful means for neutrino mass determination thanks to its very low transition energy -- $Q_{EC}$ is about 2.5 keV.
\Ho\ decays to $^{163}$Dy with a half life of about 4570 years and, because of the low transition energy, capture is only allowed from the M shell or higher.  The EC decay may be detected only through the mostly non-radiative atomic de-excitation of the Dy atom and from the Inner Bremsstrahlung (IB) radiation.

There are at least three proposed independent methods to assess the neutrino mass from the \Ho\ EC decay:  1) absolute M capture rates or M/N capture rate ratios \cite{Bennett}, 2) Inner Bremsstrahlung end-point \cite{DeRujula}, and 3) total absorption spectrum end-point \cite{Lusignoli}. 
Practically all the experimental research has focused on the atomic emissions -- photons and 
electrons -- following the EC to exploit the possibility to constrain simultaneously 
the transition $Q_{EC}$ and the neutrino mass from relative probabilities of M and N shell capture or absolute
M capture rate (method 1)) \cite{Bennett,Andersen,Baisden,Yasumi,Jonson,Hartman,Laesegaard,Gatti}. %[1,4,5,6,8,9,10,11].
The main limitation to this approach is the relatively large uncertainty on the atomic physics factors involved.
Only one experiment measured the \Ho\ IB (method 2)) but sensitivity at the end-point was impaired by background \cite{Springer}.

\begin{figure}[t!]
\begin{center}
 \includegraphics*[bb=40 20 580 420,clip,width=\linewidth]{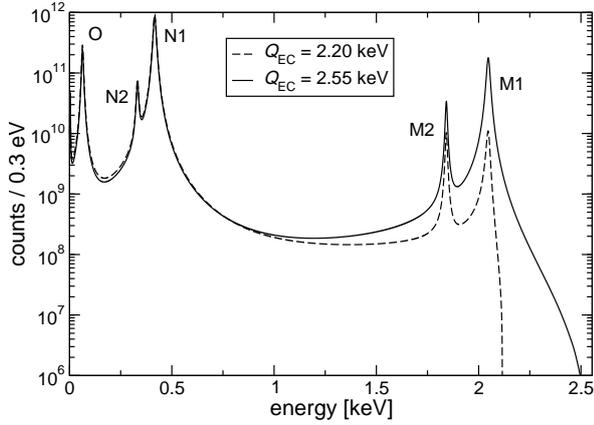}
\caption{\label{fig:ho-spe} \Ho\ total absorption spectrum calculated for an energy resolution $\Delta E_{\mathrm{FWHM}}=2$\,eV and a pile-up fraction $f_{pp}=10^{-6}$.}
\end{center}
\end{figure}

\begin{figure}[h!]
\begin{center}
% \vfill
 \includegraphics*[bb=30 20 450 420,clip,width=0.75\linewidth]{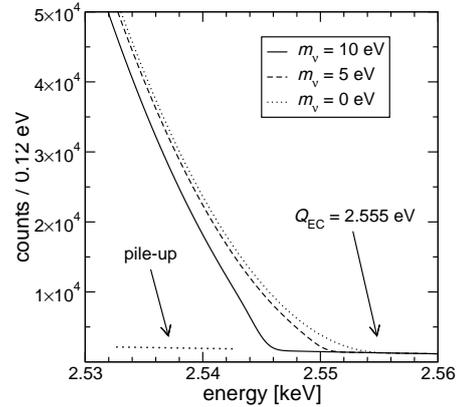}
\caption{\label{fig:ho-endpoint} Blow up of the spectrum in Fig.\,\ref{fig:ho-spe} close to the end-point.}
\end{center}
\end{figure}

On the contrary, so far there has been no experimental attempt to exploit the third method which consists in studying the end-point of the total absorption spectrum (or calorimetric spectrum, i.e. where all the energy released in the decay is measured except that carried away by the neutrino) as proposed by De Rujula and Lusignoli in \cite{Lusignoli}. 
The total absorption spectrum is made up of peaks with Breit-Wigner shapes. Positions and widths correspond to the binding energies and natural widths of the atomic levels from which the electron can be captured, while the relative intensities are given by the relative capture probabilities. The right side wings of all peaks sum up and contribute to the upper tail of the spectrum which is truncated at $Q_{EC}-m_{\nu_e}$, in analogy to what happens for nuclear \b\ decay spectra.  
In spite of the many experiments, atomic and nuclear details of the \Ho\ decay are still affected by
 unknowns which reflect in uncertainties in the expected calorimetric spectrum shape \cite{Riisager}. In particular, the various $Q_{EC}$ determinations span from 2.2 to 2.8 keV \cite{NDS} with a recommended value of about 2.555\,keV \cite{Wapstra}. Fig.\,\ref{fig:ho-spe} shows the \Ho\ EC decay calorimetric spectrum for two hypothetical $Q_{EC}$ values and for a choice of atomic parameters found in the literature. The shown spectrum  is convoluted with an instrumental energy resolution $\Delta E_{\mathrm{FWHM}}$ of about 2\,eV. The effect of a finite neutrino mass is shown in Fig.\,\ref{fig:ho-endpoint} for a $Q_{EC}$ value of 2.555\,keV. 
\begin{figure}[th]
\begin{center}
 \includegraphics*[bb=40 50 780 530,clip,width=\linewidth]{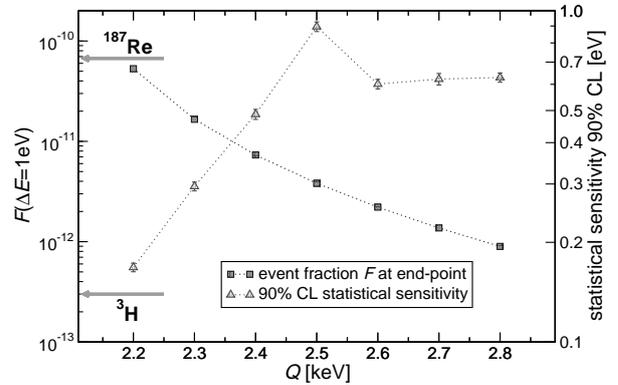}
\caption{\label{fig:ho-sens} Fraction of events in an energy interval $\Delta E = 1$\,eV below the end-point of the total absorption spectrum of \Ho\ EC decay (left Y-axis and squared) and neutrino mass statistical sensitivity for $N_{ev}=10^{14}$\,events, $f_{pp}=10^{-5}$ and $\Delta E_{\mathrm{FWHM}}=1$\,eV (right Y-axis and triangles).}
\end{center}
\end{figure}
As for \b\ decay end-point experiments, also in \Ho\ experiments the sensitivity on the neutrino mass depends on the fraction of events at the end-point: Fig.\,\ref{fig:ho-sens} shows how this fraction increases for decreasing $Q_{EC}$ and how the neutrino mass sensitivity increases accordingly. For the lowest measured $Q_{EC}$ values, the fraction of events at the end-point is about the same as for \Re. Neutrino mass sensitivity in  Fig.\,\ref{fig:ho-sens} is estimated by Montecarlo methods, as in Ref.\,\cite{Nucciotti-Sens} for \Re\ \b\ decay; the parameters of the simulations are reported in the caption. 
As discussed above, calorimetric measurements are always affected by pile-up. In the case of \Ho\ the pile-up spectrum is quite complex, being the self convolution of the spectrum shown in  Fig.\,\ref{fig:ho-spe} (see for example Fig.\,4 in \cite{Lusignoli}). The sensitivity calculated in Fig.\,\ref{fig:ho-sens} includes the effect of pile-up. The loss of sensitivity for a  $Q_{EC}$ value of about 2.5\,keV can be explained by the presence of a peak around 2.5\,keV in the pile-up spectrum. The effect of this peak can be appreciated on the full curve of Fig.\,\ref{fig:ho-endpoint} as well.

In order to assess the real potential of the total absorption approach for measuring the neutrino mass with \Ho, it is mandatory to
precisely measure the $Q_{EC}$ value and the relevant atomic parameters (i.e. natural widths and capture probabilities). 
In case these parameters turn out to be favorable, \Ho\ may then represent a very interesting alternative to \Re. Because of the relatively short half life, detectors may be realized by introducing only few \Ho\ nuclei (about $10^{11}$ for 1 decay/s) in low temperature microcalorimeters optimized for low energy X-ray spectroscopy, without any further modification.  
Given the status-of-the-art and perspectives of this class of microcalorimeters \cite{LTD13}, the \Ho\ option of the MARE project make reaching very appealing \mn\ sensitivities plausible.  Last two rows of Tab.\,\ref{tab:sens} show the
total statistics and exposures $T=N_{det}t_M$ required to reach a statistical sensitivity  $\Sigma_{90}(m_\nu)$ of 0.2 and 0.1\,eV respectively, in the fortunate case $Q_{EC}$ is 2.2\,keV.
For example, a target neutrino mass statistical sensitivity of 0.2\,eV could be expected running for just one year $1.5\times10^4$ detectors, each with a \Ho\ activity of 100\,decay/s and with energy and time resolutions of about 1\,eV and 0.1\,\mus\ respectively. The total number of implanted \Ho\ nuclei would be about $2\times10^{17}$.

\subsection{MARE-1 and MARE-2}
\label{sec:mare2}
The MARE project is subdivided into two phases. The second phase -- MARE-2 -- is the final large scale experiment with sub-electron sensitivity. The first one -- MARE-1 -- is a collection of activities with the aim of sorting out both the best isotope and 
the most suited detector technology to be used for the final experiment.

As discussed above the two competing isotopes are \Re\ and \Ho. The latter lacks the burden of 
information that have been collected for \Re\ over the years. A small scale experiment is being 
set-up by the Milano, Genova and Lisbon groups to perform a high statistics calorimetric measurement of the
\Ho\ spectrum. This will allow to determine the transition energy $Q_{EC}$, to assess the spectrum shape and therefore
to estimate the potential for a neutrino mass measurement. This experiment will use a small array of silicon thermistors 
with tin absorbers. The \Ho\ nuclei will be implanted in tin by means of a process which is currently being defined
and tested.

On the \Re\ side, the Milano group, together with the NASA/GSFC and Wisconsin groups, has developed arrays of silicon
implanted thermistors coupled to 500\,\mug\ \agre\ absorbers.
The experiment, which is presently being assembled, uses up to 8 XRS2 arrays with 36 pixels each \cite{MARE-1-Elba}.
So far, the established energy and time resolutions are about 25\,eV and 250\,\mus\
respectively. With all 288 pixels instrumented, a sensitivity better than about 4\,eV at 90\% CL would be reached in 3
years with a statistics of almost $10^{10}$ decays.
The purpose of this experiment is to investigate the systematics of \Re\ neutrino mass measurements, with a special
focus on those caused by the BEFS, the beta spectrum theoretical shape and the summing of many spectra from an array. 

More MARE-1 activities are devoted to the design of the single detector for the final MARE large scale experiment.
This mainly consists in optimizing the coupling between rhenium crystals -- or \Ho\ implanted absorbers -- and sensitive sensors like Transition Edge Sensors (TES), Metallic Magnetic Calorimeters (MMC) or Kinetic Inductance Detectors (MKID) 
In particular, groups in Genova, Miami and Lisbon are concentrating their efforts to the optimization of TES, while a group in Heidelberg focuses on MMC \cite{Heidelberg-Nu2010}.
The same groups are also developing signal multiplexing schemes. 

 \section{Conclusions}
% \label{sec:conclusions}
It is realistic to expect the MARE-1 phase to conclude with the pinning down of the best isotope and the best detector technique within 2 -- 3 years. The MARE project will then proceed to its second
phase beginning with the design and construction of one first large array. More arrays will then be added and run in different refrigerators within the collaboration.


\begin{thebibliography}{99}
\bibitem{MainzTroitsk}Ch.\,Kraus et al., \Journal{Eur.\,Phys.\,J.\,C}{40}{2005}{447}
\bibitem{KATRIN}KATRIN Design Report (2004), FZKA7090; KATRIN LoI (2001), hep-ex/0109033; T.\,Th\"ummler, this issue.
\bibitem{MIBETA-PRL} C. Arnaboldi et al., \Journal{\PRL}{91}{2003}{161802}
\bibitem{MIBETA-LTD10} M. Sisti et al., \Journal{\NIMA}{520}{2004}{125}
\bibitem{MANU-PRC} M. Galeazzi et al., \Journal{\PRC}{63}{2001}{014302}
\bibitem{MANU} F. Gatti et al., Nucl.Phys. B, 91 (2001) 293
\bibitem{proposal} the MARE proposal, http://mare.dfm.uninsubria.it
\bibitem{MARE-ltd12} A. Nucciotti, \Journal{\JLT}{151}{2008}{597}
\bibitem{Nucciotti-Sens} A. Nucciotti et al., \Journal{\APP}{34}{2010}{80}
\bibitem{Ge-Nature}F. Gatti et al.,  \Journal{Nature}{397}{1999}{137}.
\bibitem{BEFS-PRL} C. Arnaboldi et al., \Journal{\PRL}{96}{2006}{042503}
\bibitem{Foggetta-LTD12} L. Foggetta et al., \Journal{\JLT}{2008}{151}{613}
\bibitem{Gatti-LTD12} F. Gatti et al., \Journal{\JLT}{2008}{151}{603}
\bibitem{Bennett} C.L. Bennett et al., \Journal{\PLB}{107}{1981}{19}
\bibitem{DeRujula} A. De Rujula, \Journal{\NPB}{188}{1981}{414}
\bibitem{Lusignoli} A. De Rujula and M. Lusignoli, \Journal{\PLB}{118}{1982}{429}
\bibitem{Andersen} J.U. Andersen et al., \Journal{\PLB}{113}{1982}{72}
\bibitem{Baisden} P.A. Baisden et al., \Journal{\PRC}{28}{1983}{337}
\bibitem{Yasumi} S. Yasumi et al., \Journal{\PLB}{334}{1994}{229}
\bibitem{Springer} P.T. Springer et al., \Journal{\PRA}{35}{1987}{679}
\bibitem{Jonson} B. Jonson et al., \Journal{\NPA}{396}{1983}{489}
\bibitem{Hartman} F.X. Hartman et al., \Journal{\NIMA}{313}{1992}{237}
\bibitem{Laesegaard} E. Laesegaard et al., Atomic Masses and Fundamental Constants. Proceedings of the 
7th International Conference, 1984; CERN Report no. CERN-EP/84-110
\bibitem{Gatti} F. Gatti et al., \Journal{\PLB}{398}{1997}{415}
\bibitem{NDS} C.W. Reich et al., \Journal{\NDS}{111}{2010}{1211}
\bibitem{Wapstra} G. Audi and A.H. Wapstra, \Journal{\NPA}{595}{1995}{409}
\bibitem{Riisager} K. Riisager, \Journal{\JPGNP}{14}{1988}{1301}
\bibitem{LTD13} CP1185, Low Temperature Detectors LTD-13, Proceedings of the 13th International Workshop. Edited by B. Cabrera, A. Miller and B. Young, 2009 American Institute of Physics. 
\bibitem{MARE-1-Elba} A. Nucciotti et al., \Journal{\NIMA}{617}{2010}{509}
\bibitem{Heidelberg-Nu2010} P.\,Porst et al., this issue.

\end{thebibliography}
\end{document}